\documentclass[slac_one]{revtex4}
\usepackage{graphicx}
\usepackage{fancyhdr}
\usepackage{epsfig}
\begin{document}

\title{Dark Matter Candidates}

\author{Edward A. Baltz}
\affiliation{KIPAC, Stanford University, P.O.\ Box 20450, MS 29,
Stanford, CA 94309, USA}

\begin{abstract}

It is now widely accepted that most of mass--energy in the universe is
unobserved except by its gravitational effects.  Baryons make only about 4\% of
the total, with ``dark matter'' making up about 23\% and the ``dark energy''
responsible for the accelerated expansion of the universe making up the
remainder.  We focus on the dark matter, which is the primary constituent of
galaxies.  We outline the observed properties of this material, enumerating
some candidates covering 90 orders of magnitude in mass.  Finally, we argue
that the weak scale (100 GeV) is relevant to new physics, including the dark
matter problem.

\end{abstract}

\maketitle

\thispagestyle{fancy}

\section{THE DARK MATTER PROBLEM}

Cosmological parameter estimation, based on cosmic microwave background
anisotropies measured by the WMAP satellite \cite{WMAP} and based on the power
spectrum of galaxy density fluctuations measured by the SDSS collaboration
\cite{SDSS} indicates that most of matter in the universe is unobserved save by
gravity.  Approximately 4\% of the cosmological energy density is accounted for
by baryons, 23\% by the dark matter discussed in this lecture, with the
remainder being the ``dark energy'' responsible for the observed accelerated
expansion of the universe.  The dark matter is the primary component of
cosmological structures at the dwarf galaxy scale and larger.  The dark matter
and dark energy are of unknown composition.  In this lecture we discuss the
physics of dark matter, and enumerate a list of candidates, not meant to be
exhaustive.

\section{DARK MATTER PROPERTIES\label{sec:properties}}

In this section we enumerate five basic properties that dark matter is observed
to have.  The first three (that it must be non-radiating, approximately
collisionless and non-relativistic) do not otherwise place any positive
constraints on the space of possibilities, while the last two (that it must be
fluid and classical) place upper and lower bounds respectively on the mass of
the particles.

\subsection{Optically Dark (Dissipationless)}

Dark matter is not observed to shine, thus the dark matter particles must have
very weak electromagnetic interactions.  Either the electric charge and
electric and magnetic dipole moments must vanish (or be very small), or the
particles must be very heavy.  For a review of constraints on dark matter from
background light across the electromagnetic spectrum, see
Ref.~\cite{backgroundlight}.  An important consequence of this is that the dark
matter can not cool by radiating photons, and thus will not collapse to the
center of galaxies as the baryons do, by radiating their energy away
electromagnetically.  In other words, the dark matter is very nearly
dissipationless.

\subsection{Collisionless}
\label{sec:collisionless}

In addition to not interacting with ``light'' matter (to ensure that energy is
not dissipated), the dark matter must be nearly collisionless as well.  Even if
the dark matter can not radiate energy, collisions will serve to make halos
round, in contrast to data that overwhelmingly indicates triaxiality, e.g.\ in
clusters \cite{cluster_elliptical}.  The limit of this constraint is that there
can be dark matter self interactions that are important at high densities and
short distance scales, which may serve to erase small scale structure in galaxy
halos, and erase the density cusps expected at the cores of galaxies.  These
effects appear for interaction cross sections of $\sigma\sim (m/\rm GeV)
(\lambda/\rm Mpc)^{-1}$ barns \cite{spergelsteinhardt}, where $\lambda$ is the
mean free path of the particles in typical halos.  Essentially, this order of
magnitude is indicative of hadrons, and there are a number of possibilities.

\subsection{Cold}

In the early 1980's it was noted that ``cold'' dark matter explained the
observed properties of galaxies quite well \cite{CDM}.  The measured two point
correlation function of galaxies indicates that there is a large amount of
power on small scales.  If the dark matter particles have significant
velocities, then the small scale power is erased.  For this not to occur, the
particles must be sufficiently non-relativistic at the epoch of
matter-radiation equality, when the temperature of the universe was roughly 1
eV \cite{WDM}.  In practice, the constraint is that a particle species in
thermal equilibrium must have a mass larger than about 1 keV
\cite{warm_constraint}.  Of course, a non-thermal species can have a smaller
mass.  Dark matter that is not utterly cold may in fact ameliorate some of the
difficulties of the cold dark matter model \cite{WDM_bonus}; some suppression
of small scale density fluctuations may be favored.

\subsection{Fluid}

The dark matter must be sufficiently fine on galaxy scales such that the
discreteness is not detectable as yet.  There are two basic effects to be
concerned with.  The first is that the granularity of dark matter provides a
time dependent gravitational potential, which may disrupt bound systems.  For
particles of $10^6\,M_\odot$, this effect would heat the galactic disk to
observable levels, while globular clusters would be disrupted for smaller
masses, around $10^3\,M_\odot$ \cite{discreteheating}.  Second, the
discreteness of dark matter introduces Poisson noise in the power spectrum of
density fluctuations, which conflicts with observations of the Ly$\alpha$
forest if the particles are more massive than roughly $m\agt10^4\,M_\odot$
\cite{discretepoisson}.  We thus find that an upper bound on the mass of
particles of roughly $m\alt10^{3-4}\,M_\odot\sim10^{70-71}$ eV.

Gravitational microlensing places further constraints on dark matter in the
solar mass range.  Searches for lenses by the MACHO and EROS collaborations
have failed to find enough microlensing events to explain the full dark halo of
the Milky Way \cite{microlensing}, though MACHO reports that 20\% of the Milky
Way halo consists of objects of roughly 0.3 $M_\odot$.  The constraints are
that objects between $10^{-7}\,M_\odot$ and $10\,M_\odot$ do not make up more
than about 20\% of the dark matter in the Milky Way halo.  Microlensing surveys
are not able to place constraints much outside this mass range due to the
limitations of microlensing event timescales.

\subsection{Classical}

Dark matter must behave sufficiently classically to be confined on galaxy
scales.  Bounds can be placed on the masses of both bosons and fermions, based
only on the observed properties of galaxies: namely galaxy densities must reach
of order GeV cm$^{-3}$, their velocity dispersions are of order 100 km
s$^{-1}$, and their sizes are of order kpc.

\subsubsection{Bosons}
\label{sec:fuzzy}

If dark matter consists of bosons, their quantum nature is manifest only if
their mass is exceedingly small.  In order to form galaxies, dark matter
particles must be confined on kpc scales.  With typical galactic velocities,
the de Broglie wavelength of the particles is $\lambda\sim ({\rm eV}/\,m)$ mm.
Setting this wavelength to 1 kpc, we find that particles with mass $m\sim
10^{-22}$ eV are barely confined.  This mass denotes the lower end of the
possible range for dark matter.  In fact it was proposed as a mechanism to
erase the small scale power in the galaxy power spectrum \cite{fuzzy}, called
``fuzzy'' cold dark matter.  This also disallows the central density cusp found
in numerical simulations of galactic dark halos.

\subsubsection{Fermions}
The lower bound on the mass of a fermionic dark matter particle is much more
stringent than that for bosons.  As first pointed out in
Ref.~\cite{tremainegunn}, the phase space density of fermions has a maximum
value $f=g h^{-3}$, where $g$ is the number of internal degrees of freedom, and
in fact for a relativistic gas of fermions in thermal equilibrium it is half
this value.  In a galactic halo, the maximum phase space density assuming a
Maxwellian velocity distribution is simply $f=\rho/[m^4(2\pi\sigma^2)^{3/2}]$.
Pauli blocking then enforces $m^4>\rho h^3/[g(2\pi\sigma^2)^{3/2}]$.  Taking a
Milky Way type galaxy, where $\rho>1$ GeV cm$^{-3}$ is required in the center
and $\sigma=150$ km s$^{-1}$, and assuming $g=2$, we find $m\,\agt$ 25 eV for
fermions.

\section{COSMOLOGICAL RELIC DENSITY OF PARTICLES}
\label{sec:relicdensity}

The dark matter problem is a significant challenge for the particle physics
community in that any candidate must satisfy stringent constraints.  From a
particle physics point of view, the properties discussed in
section~\ref{sec:properties} are not hard to arrange, but two implicit
properties are: namely that the candidate have a lifetime much longer than the
Hubble time $\sim$ 10 Gyr, and also that its cosmological density is compatible
with the observed dark matter density.  The current measured value of the dark
matter density is $\Omega_{\rm CDM} h^2=0.135^{+0.008}_{-0.009}$ \cite{WMAP},
where $\Omega$ is the density relative to the critical density
$\rho_c=3H_0^2/(8\pi G)$, and $h$ $(=0.71^{+0.04}_{-0.03})$ is the scaled
Hubble constant $H_0=100\,h$ km s$^{-1}$ Mpc$^{-1}$.  In particle physics
units, $\rho_c\approx 10.5\,h^2$ keV cm$^{-3}$.  Accounting for the possibility
of a significant neutrino density (see section~\ref{sec:neutrinos}), we take a
95\% confidence region $0.087<\Omega_{\rm CDM}h^2<0.129$.

The simplest possibility would be for a candidate to have the correct density
from thermal processes alone.  Computing the relic density proceeds as follows
\cite{KT}.  The species is assumed to be in thermal equilibrium at early enough
times, i.e.\ when the universe was essentially a radiation dominated plasma.
The production and annihilation rates are tracked as the universe expands, and
when these rates become longer than the (current) Hubble time, the species is
said to be ``frozen out'' with a comoving density that no longer changes.  This
statement is expressed in the Boltzmann equation,
\begin{equation}
\frac{dn}{dt}+3Hn=-\langle\sigma v\rangle\left(n^2-n_{\rm eq}^2\right),
\end{equation}
where $n$ and $n_{\rm eq}$ are the particle density and its equilibrium value,
$H$ is the Hubble parameter and $\langle\sigma v\rangle$ is annihilation cross
section, appropriately thermally averaged \cite{gondologelmini}.  We
immediately notice that in the absence of interactions, the number density
evolves exactly as expected, $n\propto a^{-3}$, where $a$ is the cosmic scale
factor and $H=\dot{a}/a$.  The relativistic gas has a temperature evolving as
$T\propto a^{-1}$, thus $n\propto T^3$.  The details of freeze out now depend
on whether the relic is relativistic or not.

\subsection{Relativistic Thermal Freeze-out}

For relativistic particles, $n_{\rm eq}\propto T^3$, thus a species that has
frozen out will still track the equilibrium density.  The relic density of such
particles is thus insensitive to the details of freeze-out.  The only relevant
parameter is the effective number of degrees of freedom at freeze-out,
$g_\star$.  This can change as a function of temperature, thus the density of a
frozen out species need not track its equilibrium density, but the scaling with
temperature remains as long as the particles are relativistic.  The well known
example of neutrinos freeze out with $g_\star\approx 10.75$, prior to the
freeze-out of electrons and positrons (which lowers $g_\star$).  Thus the
(massless) neutrino temperature is lower than the photon temperature in the
present universe.  The relic density of any relativistic species is given by
\begin{equation}
\Omega h^2\approx\frac{m}{100\,\rm eV}\left(\frac{g_\star}{10}\right)^{-1}.
\end{equation}

\subsection{Non-Relativistic Thermal Freeze-out}

Non relativistic particles have an exponentially suppressed equilibrium
density, $n_{\rm eq}\propto (mT)^{3/2}\exp(-m/T)$.  A frozen out species would
thus have a density much larger than its equilibrium density.  We can simply
derive the scaling of relic density with annihilation cross section as follows.
In a radiation dominated universe, $H\propto T^2$ (since $H^2\propto\rho$
according to the Friedman equation).  The freeze-out condition can then be
expressed simply as $H(T_{\rm fr})\sim \sigma n_{\rm eq}(T_{\rm fr})$.  We find
that $T_{\rm fr}\propto m$, with only logarithmic corrections (this is due to
the exponential Boltzmann factor).  We can then derive the current mass density
as $\rho\sim m n_{\rm eq}(T_{\rm fr}) (T/T_{\rm fr})^3\sim T^3 (m/T_{\rm
fr})/\sigma$.  Thus we find the crucial result that $\rho\propto 1/\sigma$,
with only logarithmic dependence on mass since $m/T_{\rm fr}$ is roughly
constant.  This means that smaller annihilation cross sections yield larger
relic densities, which makes sense, as less efficient annihilations should
allow more particles to survive.  Taking the simple scaling that $\sigma\propto
m^{-2}$, we find the approximate relic density
\begin{equation}
\Omega h^2\sim\left(\frac{m}{\rm TeV}\right)^2.
\end{equation}
This is a very interesting result in that a stable particle at the weak
interaction scale of several hundred GeV would give the proper relic density to
be dark matter.  Any Weakly Interacting Massive Particle (WIMP) might thus be a
compelling dark matter candidate.

\subsection{Pedagogical Example}

\begin{figure}[h]
\epsfig{file=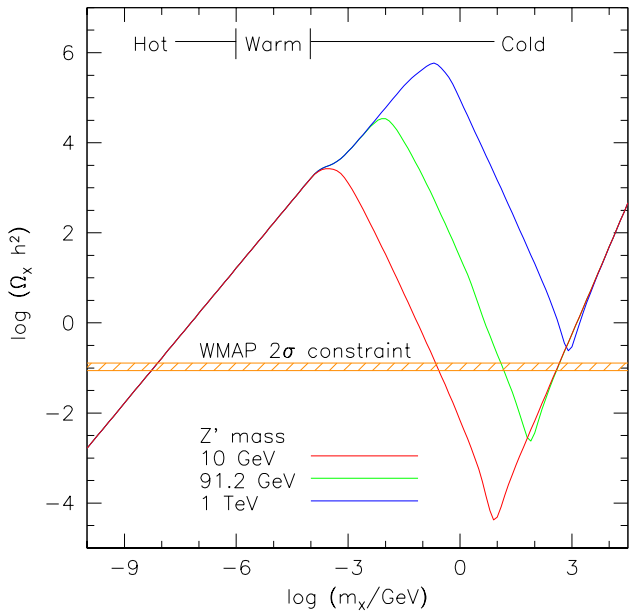,width=0.5\textwidth}
\raisebox{1.9in}{\parbox{0.49\textwidth}{\caption{Relic density of a species
freezing out in thermal equilibrium.  The interaction cross section is of weak
interaction strength.  For masses below $\sim$ MeV, the particles are
relativistic at freeze-out, above this mass they are non-relativistic.  Above a
mass of $\sim$ 100 GeV, the interactions look ``electromagnetic'' as the gauge
boson becomes light relative to the particle mass.  The three regimes of cold,
warm, and hot dark matter are illustrated, along with the WMAP $2\sigma$
constraint on dark matter density.\label{fig:thermal}}}}
\end{figure}

We can illustrate both the relativistic and non-relativistic freeze-out regimes
with a single toy model as follows.  Taking a Dirac fermion annihilating
through a wide $Z'$ boson, the annihilation cross section is
\begin{equation}
\sigma v=
\left(\frac{g^4}{64\pi\cos^4\theta_W}\right)\,\frac{m^2}{(s-m_{Z'}^2)^2+
m_{Z'}^4}.
\end{equation}
In figure~\ref{fig:thermal}, we plot the relic density of this particle for a
wide range of masses, as calculated by the DarkSUSY code \cite{darksusy}.
Three regimes are evident.  For masses below 1 MeV, the freeze out is
relativistic, and the relic density is proportional to mass.  Above 1 MeV, but
below $m_{Z'}$ the interaction is like low energy weak interactions, with
$\sigma \propto E^2=m^2$, thus the relic density is proportional to the inverse
square of the mass.  For masses larger than the $Z'$, the usual behavior of
$\sigma\propto m^{-2}$ is recovered, with relic density proportional to the
square of the mass.  In a parallel set of three regimes, the boundaries between
hot, warm, and cold dark matter are approximately illustrated.  Comparing with
the known value of the relic density, the hot and cold possibilities appear at
10 eV and 1 TeV respectively.

%\vspace{-.5in}

\section{DARK MATTER CANDIDATES}

In this section we outline some of the proposals for dark matter.  While this
list is by no means exhaustive, we will attempt to cover the range of
possibilities that have been considered at least qualitatively.  We will
proceed very roughly from the smallest mass candidates at $10^{-22}$ eV (the
fuzzy cold dark matter discussed previously) to the largest at $10^6
M_\odot\approx10^{72}$ eV.  This range covers more than 90 orders of magnitude
in mass; unfortunately observations do not place stronger bounds on possible
candidates.  However, the weak scale (100 GeV) is quite interesting for the
dark matter problem, as we will describe.

\subsection{``Fuzzy'' Cold Dark Matter}

This model was discussed in section~\ref{sec:fuzzy}.

\subsection{Chaplygin Gas}

The Chaplygin gas is an attempt to unify dark matter and dark energy in a
single fluid with an equation of state $p=-A/\rho$.  Generalizations of this
have also been considered \cite{chaplygin}.  This type of fluid can arise in
certain string-inspired models involving $d-$branes.  The transition between
the dust-like universe at moderate redshift to the accelerating universe today
occurs naturally in these models.

\subsection{Axions}
\label{sec:axion}

Strong interactions naively violate $CP$ with order unity strength (the
coefficient $\theta$ of the $F\tilde F$ term in the QCD Lagrangian is not
obviously small), while experimentally the violation must be smaller than
$\theta\sim 10^{-9}$.  Promoting the parameter to a dynamical field solves this
problem, at the expense of a new light pseudoscalar boson, the axion
\cite{axion}.  The cosmology and astrophysics of these particles has been
studied at length, beginning with Ref.~\cite{axioncosm} and reviewed in
Refs.~\cite{axionreview}.  To summarize, there is an allowed mass window of
interest to dark matter, with $\mu$eV $\alt m\alt$ meV.  The lower bound comes
the fact that the relic density of axions $\Omega_a h^2\simeq(4\,\mu{\rm
eV}/m_a)^{7/6}(200\,{\rm MeV}/\Lambda_{\rm QCD})^{3/4}$, while the upper bound
comes from the fact that the axion interactions must be weak enough that stars
cool via photons and not axions.  Axion production is non-thermal, thus they
are cold, even though their masses are far below the $\sim$ keV limit for warm
dark matter.  Early on it was pointed out that axions could be detected by
resonant axion-photon conversion in a magnetic field \cite{sikivie}, an
experimental program that has been undertaken with some success
\cite{axionexpt}.  Axion dark matter is particularly interesting from an
experimental point of view in that the velocity distribution of the local dark
can be measured with high accuracy.  A record of the process of virialization
would be evident, of considerable interest to studies of galaxy formation.

\subsection{Neutrinos}
\label{sec:neutrinos}

We now know that neutrinos are one component of the cosmological dark matter.
Neutrino oscillation experiments, both for atmospheric neutrinos
\cite{superKneut} and for solar neutrinos \cite{SNO}, have now conclusively
shown that at least two of the neutrinos in the Standard Model are massive.
Measurements of the endpoint spectrum of tritium beta decay place a 95\%
confidence upper bound on the electron neutrino mass, $m_{\nu_e}\alt2.5$ eV
\cite{neutrinomass}, thus bounding all three neutrino species as the mass
splittings measured by the oscillation experiments are both much smaller than
this.  The cosmological density of light neutrinos ($m_\nu\alt$ MeV) is known
to be $\Omega_\nu h^2=\sum m_\nu/(94.0\,\rm eV)$ \cite{cowsikmcclelland}.  The
WMAP bound \cite{WMAP} on this quantity is $\Omega_\nu
h^2<0.0076\rightarrow\sum m_\nu\alt0.7$ eV, thus the cosmological bound is more
stringent than the laboratory one ($\sum m_\nu\alt7.5$ eV).  The measured
oscillation parameters enforce that $\sum m_\nu\agt50$ meV, and thus
$0.0005<\Omega_\nu h^2<0.0076$, a known constituent of dark matter as even the
bottom of this mass range would be non-relativistic today.  However, no fermion
in this mass range can be a significant constituent of galaxies due simply to
the Pauli exclusion principle discussed above.  In addition to not being
abundant enough, these Standard Model neutrinos are ``hot'' and thus unsuitable
to explain the properties of galaxies.

Very massive neutrinos have also been considered as dark matter candidates,
first in Ref.~\cite{leeweinberg}.  We now know that none of the Standard Model
neutrinos are suitable, but their right-handed counterparts might be
acceptable.  Active neutrinos in a 4th generation would be acceptable except
for the fact that they would have been detected in sensitive underground
experiments by now.

\subsection{MeV Dark Matter}

The INTEGRAL satellite has observed an excess of 511 keV positron annihilation
emission toward the galactic center, which is difficult to reconcile with
astrophysical sources.  It has been suggested that this emission might be due
to $e^+e^-$ pairs that are the annihilation products of dark matter particles
in the MeV range \cite{MeVDM}.  Their mass must be low to exclude hadronic
final states, which would overproduce photons from e.g.\ $\pi^0$ decays, and
also so that the leptons are not so energetic that they escape the galactic
center region.  However, the particle physics motivation for this scenario is
weaker than for other candidates.

\subsection{Supersymmetry}

Our current understanding of physics includes no fundamental relationship
between bosons and fermions.  Supersymmetry was proposed as just such a
relationship, first on the 2-d worldsheet for string theory
\cite{worldsheetSUSY}, and slightly later for standard 4-d spacetime
\cite{spacetimeSUSY}.  Supersymmetric theories put bosons and fermions in
common multiplets; for most such theories then the corresponding states must
have similar masses.  It is thus obvious that if supersymmetry is a valid
symmetry, it is badly broken in the low energy world in which we live.  These
theories admit the so-called non-renormalization theorem, which remains valid
if the supersymmetry is broken ``softly''.  Very roughly stated, the only
infinite renormalizations are of wavefunction type, and those are only
logarithmically divergent.  The quadratically divergent corrections in the
Standard Model are absent.

The Minimal Supersymmetric Standard Model (MSSM, for early reviews see
Refs.~\cite{MSSM}) was proposed as a model with broken supersymmetry.  Each
Standard Model state has a superpartner with spin differing by $\hbar/2$:
matter fermions (quarks, leptons) have scalar partners, gauge bosons have spin
1/2 partners, and Higgs bosons (two $SU(2)$ doublets are required) also have
spin 1/2 partners.  The soft supersymmetry breaking in this model allows for
the masses of superpartners to differ from their Standard Model counterparts.
The most general gauge invariant Lagrangian involving these fields includes
terms that violate either baryon or lepton number, e.g.\ a term allowing scalar
up-type squarks to decay directly to a down quark and charged lepton.  If both
baryon and lepton number violating interactions are included, proton decay
occurs at the weak scale.  The usual solution is to impose R-parity
conservation, where superpartners have negative R-parity and thus must interact
in pairs.  This means that the lightest superpartner (LSP) is absolutely
stable.  Early on, it was realized that the LSP (first considered to be the
gravitino) was thus a dark matter candidate \cite{Rparity}.

\subsubsection{Gravitinos}

The superpartner of of the graviton, the spin 3/2 gravitino, was the first SUSY
particle considered for the dark matter problem \cite{Rparity}.  In models
where the gravitino is the LSP, it is often quite light (keV), and would thus
be warm dark matter.  In cosmology, the overproduction of gravitinos is
somewhat problematic, though not insurmountably so \cite{gravitino}.
Gravitinos at the weak scale, whose relic density would be obtained through the
decays of the next lightest superpartner, are also an interesting possibility
\cite{superwimp}.

\subsubsection{Neutralinos}

The favored supersymmetric dark matter candidate is the lightest neutralino;
these are the four spin 1/2 Majorana fermion superpartners of the neutral gauge
and Higgs bosons (usually denoted $\chi^0_{1-4}$) \cite{neutralino}.
Similarly, there are two charged Dirac fermion superpartners of charged gauge
and Higgs bosons, the charginos $\chi^\pm_{1-2}$.  The successes of
supersymmetric models depend on the fact that the gauginos have weak-scale
masses, thus the proper relic density for any stable states comes essentially
for free.  We focus on neutralinos as dark matter in
section~\ref{sec:neutralino}.

\subsubsection{Sneutrinos}

The scalar partners of the neutrinos are possible dark matter candidates, but
they are disfavored for two reasons.  First, they annihilate quite efficiently,
requiring masses above 500 GeV to provide a relic density consistent with dark
matter.  Second, their elastic scattering cross sections on nuclei are quite
large (on the order of femtobarns), to the point that they would be easily
detectable in current experiments \cite{sneutrino}.

\subsubsection{Axinos}

If the axions proposed to solve the strong $CP$ problem exist (see
section~\ref{sec:axion}) and in addition supersymmetry is valid, the axion will
naturally have a spin 1/2 partner, the axino.  Depending on the conditions in
the early universe, these might be either warm or cold dark matter
\cite{axino}.

\subsubsection{Q-balls}

Supersymmetric theories typically permit non-topological solitons, dubbed
Q-balls.  These carry baryon and/or lepton number, and can be absolutely stable
if large enough.  They are an interesting dark matter candidate, and may even
have relatively strong self-interactions at the level discussed in
section~\ref{sec:collisionless} \cite{Qball}.

\subsubsection{Split SUSY}

Recently it has been noted that the successes of supersymmetric models hinge on
the fact that the gauginos have masses at the weak scale, while having the
scalar superpartners at the weak scale is somewhat problematic.  If the Higgs
fine-tuning problem is simply ignored (as the cosmological constant fine-tuning
often is), the scalar superpartners can be made very heavy, keeping light
gauginos and Higgsinos \cite{splitSUSY}.  This scenario has an interesting
phenomenology, both for collider signatures and for dark matter
\cite{splitSUSYimp}.

\subsection{Universal Extra Dimensions}

If our four-dimensional spacetime is embedded in a higher dimensional space,
excitations of Standard Model states along the orthogonal dimensions (called
Kaluza-Klein excitations) may be viable dark matter candidates \cite{UED}.  For
example, the first excitation of the $B$ boson (associated with the $U(1)_Y$
hypercharge gauge group of the Standard Model) has been considered.  This is
similar to supersymmetry in that known particles have partners, but the
partners' spin does not differ.  This scenario has been dubbed ``bosonic
supersymmetry''.  The stability of the lightest of the KK excitations can be
arranged by a parity symmetry, and masses around 1 TeV provide reasonable relic
densities from thermal freeze-out, the same as with supersymmetric models.

\subsection{Branons}

String theory naturally contains objects of many different dimensions, called
branes.  These would naturally have fluctuations characterizable as particles,
the so-called branons.  These fluctuations can be made into suitable cold dark
matter candidates, both thermal and non-thermal \cite{branon}.

\subsection{Mirror Matter}

The (modern) concept of a mirror world is an old one, extending back to the
non-conservation of parity in weak interactions.  In a mirror scenario, the
dark matter could just be ordinary matter in the mirror world, where the only
communication is gravitational.  This scenario can be constructed in a
braneworld context, where our universe and a mirror universe are two branes in
a higher dimensional space.  For more complete discussions of this scenario,
see e.g.~\cite{mirror}.

\subsection{WIMPzillas}

At the end of inflation, gravitational interactions alone can produce copious
particles.  For mass scales of $10^{13}$ GeV, these particles, if stable, could
even account for the dark matter \cite{wimpzilla}.  In addition, such particles
might decay with a lifetime much longer than the age of the universe, providing
a source of ultra high energy cosmic rays -- this is the so-called ``top-down''
scenario for UHECR production.

\subsection{Primordial Black Holes}

Under the right conditions, primordial black holes can form in the early
universe, e.g.\ see Ref.~\cite{PBH}.  Production is enhanced during periods
where the equation of state softens ($p<\rho/3$), such as during a first order
phase transition.  This is easy to understand: if the pressure support
lessens, objects collapse more easily.  The last such phase transition in the
universe is the quark-hadron phase transition, at a temperature $T\sim100$ MeV.
The mass contained in the horizon at this epoch is very roughly a solar mass.

\section{FOCUS ON NEUTRALINOS}
\label{sec:neutralino}

The lightest neutralino is the dark matter candidate that has endured the most
scrutiny.  In this section we elaborate on the discussion of the MSSM, and
discuss the prospects for discovering neutralinos astrophysically.  The MSSM
starts with the Standard Model field content, adds one additional $SU(2)$
doublet of Higgs bosons (with the opposite $U(1)_Y$ charge), and then adds the
superpartners of each of these states.  The Lagrangian includes all SUSY
conserving terms that respect R-parity, and SUSY-breaking soft terms.  More
than 100 new parameters are required.  One parameter of the Standard Model is
actually removed, the quartic self-coupling of the Higgs boson.  These
couplings appear in the MSSM, but with the quartic coupling $\lambda_{\rm SM}$
replaced by terms in the gauge couplings $g^2$ and $g'^2$.  Clearly, a theory
with 100+ free parameters is not very predictive.  To make some connection to
reality, a simpler model is required.

\subsection{Minimal Supergravity}

The best studied model of low-energy supersymmetry uses gravitational effects
to break supersymmetry, called Minimal Supergravity (mSUGRA) \cite{msugra}.
The model has four parameters, plus one sign, beyond the Standard Model ones
(and again, $\lambda_{\rm SM}$ is absent).  These continuous parameters are
$m_0$, the universal scalar mass at the GUT scale, $m_{1/2}$, the universal
GUT-scale gaugino mass, $A_0$, the GUT-scale trilinear coupling, and
$\tan\beta$, the weak-scale ratio of the vacuum expectation values of the two
scalar neutral Higgs fields.  The Higgsino mass parameter $|\mu|^2$ is
determined by enforced radiative electroweak symmetry breaking, but the phase
of $\mu$ is undetermined: it is chosen to be $\pm 1$ to conserve $CP$.

In studying this model, a set of parameters is evolved according to the
renormalization group equations (RGEs), checking that the gauge couplings
unify, and electroweak symmetry breaking occurs.  There are a number of
publicly available software packages that solve RGEs: we use ISAJET
\cite{isajet}.  Further tests against accelerator data (e.g.\ sparticle masses,
BR($b\rightarrow s\gamma$), etc.) are performed.  Finally, parameter sets where
the LSP is not a neutralino are discarded.

We can roughly understand the weak-scale parameters as follows.  As the
gaugino masses are unified at the GUT scale, their weak-scale masses are
related to the ratios of the gauge coupling constants at the weak scale:
strong, weak, hypercharge gaugino masses have $M_3:M_2:M_1\sim6:2:1$, namely
the B-ino is the lightest gaugino.  Gauge couplings run scalar masses higher,
while Yukawa couplings run scalar masses lower: squarks are heavy, sleptons are
light, with $\tilde{u}_L$ usually heaviest and $\tilde{\tau}_R$ usually
lightest.

To connect these models to cosmology, we must calculate the relic density of
neutralinos.  We use the publicly available DarkSUSY code for this purpose
\cite{darksusy}.  The code actually calculates the relic density of all
neutralinos, charginos and sfermions; if the mass splittings are small, the
``coannihilation'' effects can be crucial.  Given the current experimental
limits, three regions of mSUGRA parameter space stand out as giving a relic
density compatible with cosmology: stau coannihilations, where the
$\chi^0_1-\tilde{\tau}_R$ splitting is small, the focus point region where the
$\chi^0_1-\chi^0_2-\chi^\pm_1$ splittings are small, and the region (at large
$\tan\beta$) where the annihilation through the pseudoscalar Higgs $A^0$ is
resonant, \mbox{$\chi\chi\rightarrow A^0\rightarrow f\bar{f}$}.

\subsection{MCMC Scanning}

\begin{figure}[h]
\epsfig{file=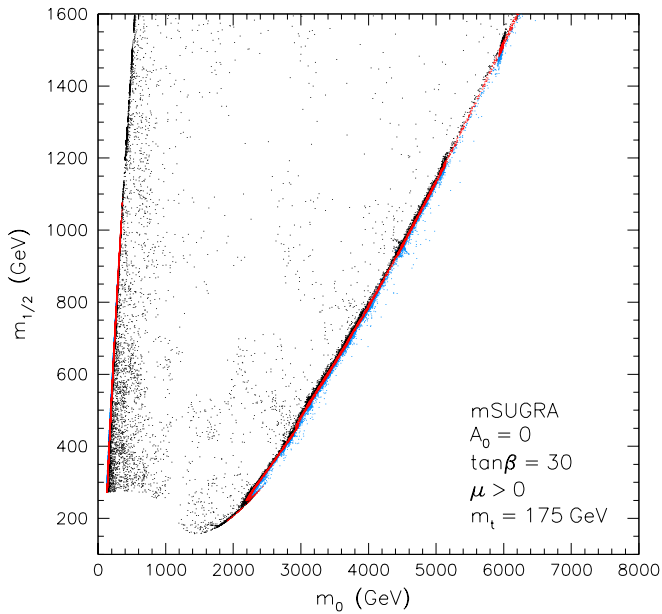,width=0.5\textwidth}
\raisebox{1.9in}{\parbox{0.49\textwidth}{\caption{MCMC scan in 2 dimensions.
We have fixed $A_0=0$, $\tan\beta=30$, $m_t=175$ GeV.  Black points illustrate
models with relic density too high, red points pass the 95\% confidence WMAP
cut, and blue points fall below (thus they are acceptable, but not as a full
solution for the dark matter problem.  The stau coannihilation region clearly
appears at the left edge, while the focus point region (chargino
coannihilation) is clear at the right edge.  At larger $\tan\beta$, the Higgs
resonance region would appear.  The stau coannihilation region ends at
$m_{1/2}\sim1100$ GeV, while the focus point extends at least to
$m_{1/2}\sim20$ TeV (10 TeV neutralinos).
\label{fig:MCMC}}}}
\end{figure}

The regions of mSUGRA parameter space compatible with cosmology are quite
narrow, and difficult to find in general.  Grid searches of the parameter space
are $\sim1\%$ efficient at finding models passing the $95\%$ confidence cut
mentioned in section~\ref{sec:relicdensity}.  However, we can use the relic
density of models to guide the search through parameter space, using the
so-called Markov Chain Monte Carlo (MCMC) \cite{mcmc}.  The basic idea is to
use a likelihood function ${\cal L}=e^{-\chi^2/2}$ (here taken as the WMAP
likelihood function for relic density) calculated at the current point in
parameter space.  The likelihood is calculated at a proposed point, and the
``chain'' advances if the likelihood is higher.  With lower likelihood, the
chain advances with probability $\cal L_{\rm new} / \cal L_{\rm old}$.  In this
way chains of models can be constructed that efficiently sample the interesting
regions.  A 2 dimensional example is illustrated in figure~\ref{fig:MCMC}.  In
fact we have explored the full 4 dimensional mSUGRA parameter space, finding
efficiencies with MCMC of 20\%-25\% (a {\em huge} improvement).  We scanned 2.4
million models in total, finding 500 thousand passing the WMAP cut.

\subsection{Direct Detection}

\begin{figure}[h]
\epsfig{file=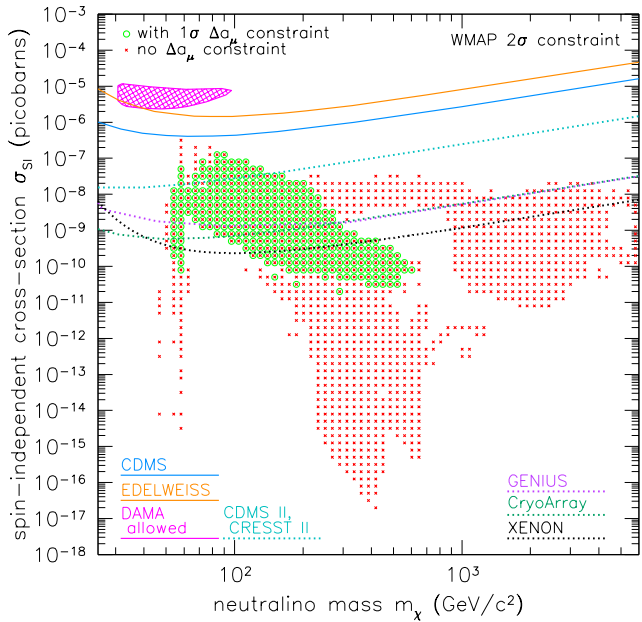,width=0.5\textwidth}
\raisebox{1.9in}{\parbox{0.49\textwidth}{\caption{Direct detection of
neutralinos.  The spin-independent elastic scattering cross section of
neutralinos on protons is plotted against neutralino mass.  Red crosses
indicate the mSUGRA parameter space, while green circles indicate those models
that could make a significant contribution to the $g-2$ of the muon, which may
be experimentally favored \cite{gminus2}.  Current limits from the EDELWEISS
and CDMS experiments are plotted \cite{ddcurrent}, along with the projected
sensitivities of several proposed experiments \cite{ddproposed}.  The region
consistent with the DAMA annual modulation claim is illustrated as well
\cite{dama}; this region is inconsistent with the negative results of other
experiments.
\label{fig:DD}}}}
\end{figure}

\begin{figure}[h]
\epsfig{file=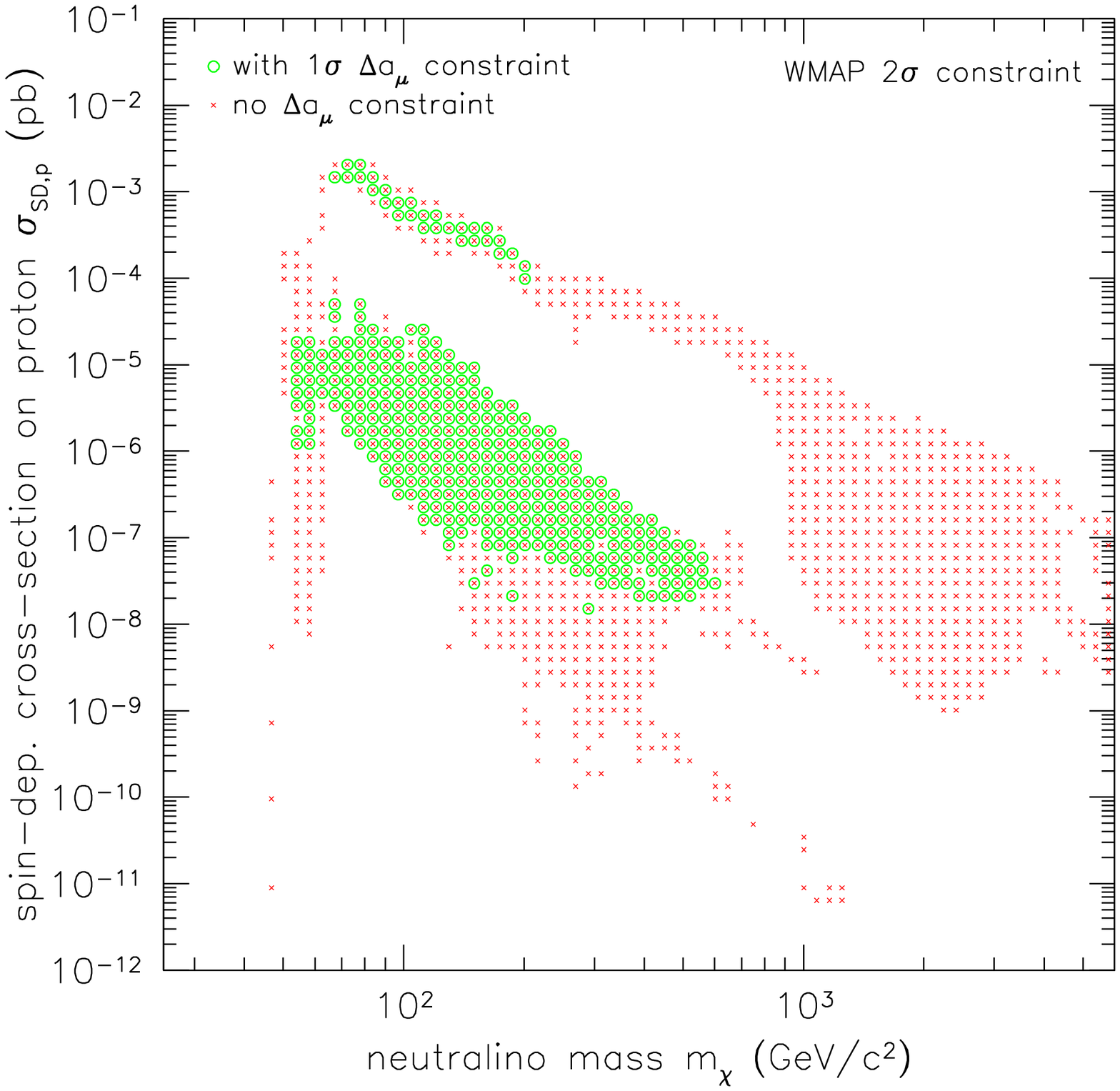,width=0.49\textwidth}
\epsfig{file=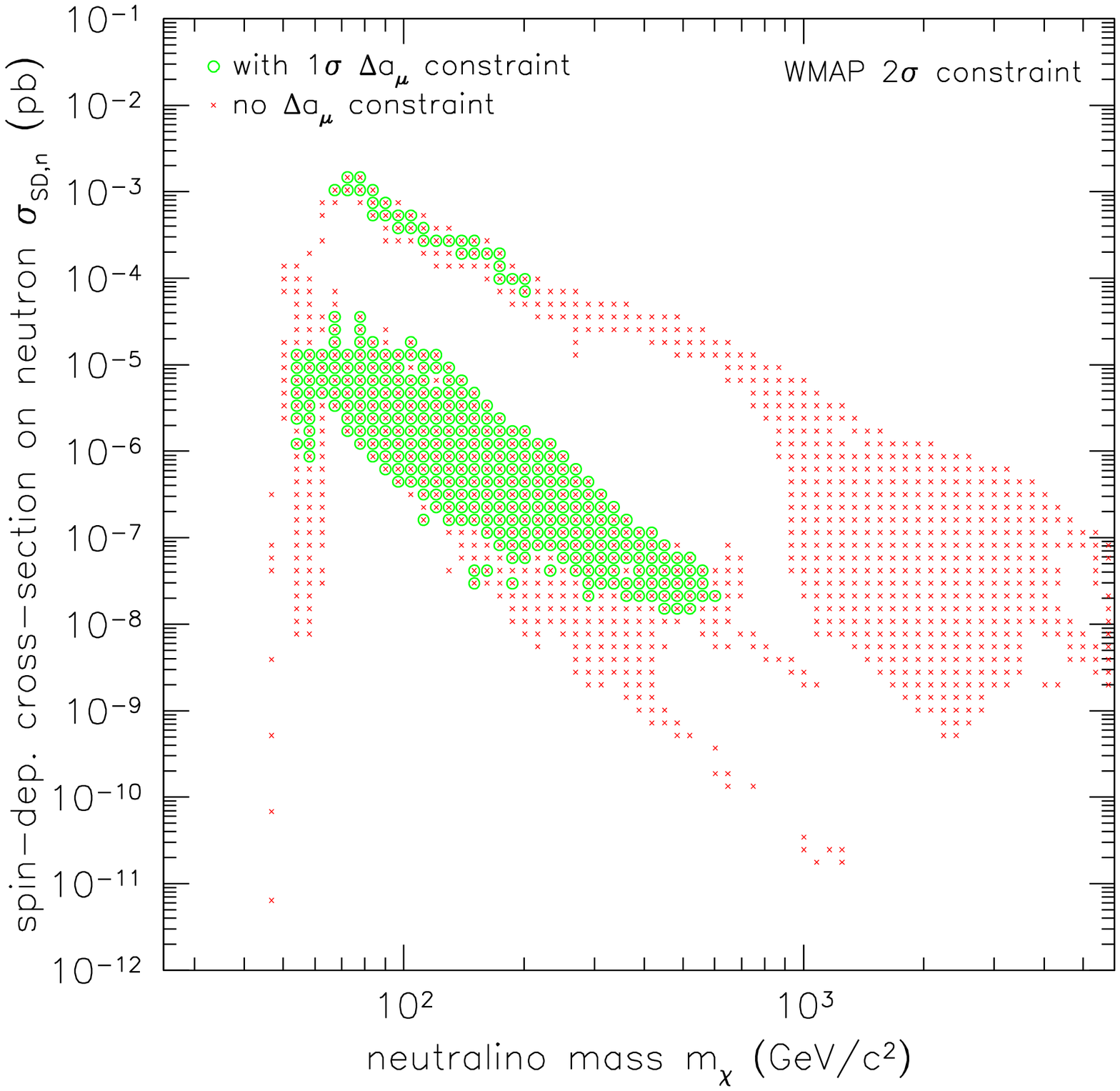,width=0.49\textwidth}
\caption{Direct detection of neutralinos.  The spin-dependent elastic
scattering cross sections of neutralinos on protons and neutrons are plotted
against neutralino mass.  Symbols are as in figure~\ref{fig:DD}.
\label{fig:DDSD}}
\end{figure}

On of the most promising experimental techniques for discovering neutralinos is
``direct detection''.  The idea is that the neutralino will scatter on a
nucleus, depositing energy that can be detected in a number of ways
\cite{goodmanwitten}.  The scatterings are very rare (the current limits are of
the order of $\sigma\sim10^{-42}$ cm$^2$), and they deposit very little energy
(typically $\sim 10$ keV), so the experimental challenge is considerable.  In
figure~\ref{fig:DD}, we plot all of the mSUGRA models consistent with solving
the dark matter problem.  The axes are neutralino mass and spin-independent
neutralino-nucleon elastic scattering cross section.  The spin independent
scattering amplitudes essentially couple to the mass of the nucleon, and add in
quadrature due to coherent effects (the 10 keV exchanged is long wavelength
compared to a nucleus).  Current experimental efforts are just starting to
probe the mSUGRA region, with several orders of magnitude improvement being
quite attainable in the near future.  Spin-dependent interactions are also
present, but they are experimentally more difficult to explore, requiring odd-A
targets (though the cross sections can be quite large).  In
figure~\ref{fig:DDSD} we plot the spin-dependent cross sections on protons and
neutrons.

\subsection{Indirect Detection}

Here, we will briefly mention the indirect techniques for neutralino searches.
The first relies on the fact that massive objects such as the Earth or Sun will
collect neutralinos due to elastic scattering into bound gravitational orbits.
The particles thereafter become concentrated near the center of such objects,
and annihilate more rapidly.  The neutrinos from the annihilation products can
be searched for in neutrino telescopes \cite{earthsun}.

Neutralino annihilations in the galactic halo proceed at a very small rate, but
the annihilation products are very energetic and might thus be detectable.
These are (focusing on antiparticles) gamma rays, antiprotons, positrons, and
antideuterons.  Monochromatic gamma rays are particularly interesting; the
$\chi\chi\rightarrow\gamma\gamma,\,Z^0\gamma$ branching ratios are typically
quite small, but the signal is unmistakable and has no astrophysical
background.

\section{CAN WE LIVE WITHOUT DARK MATTER?}

Observed matter can not account for the gravitational potentials of objects at
the dwarf galaxy scale and larger, thus dark matter was hypothesized.  An
alternative possibility is that gravity weakens less quickly at large
distances, or as proposed by Milgrom \cite{oldMOND}, at small acceleration
scales.  The latter idea is known as MOND, for MOdified Newtonian Dynamics.
Basically, imposing a {\em minimum} acceleration scale $a_0\sim cH_0$ turns out
to explain spiral galaxies quite well.  If this MOND scale $a_0$ is allowed to
run, clusters might be explained as well.  Because gravitational lensing mass
estimates agree with dynamical estimates, the MOND acceleration scale must
apply to photons as well.  A relativistic theory is needed in any case to
construct a consistent cosmology.  A number of attempts have been made, each
with fatal inconsistencies \cite{relMOND}.  Quite recently, Bekenstein has
proposed a tensor-vector-scalar theory (TeVeS) that is successful in
reproducing MOND in the proper limit even for photons, and respecting the
classical tests of general relativity \cite{newMOND}.  Whether a consistent
cosmology can be constructed with the theory remains to be seen.  Note that
even this is a ``dark matter'' theory, in that the structure of galaxies is
still explained by the introduction of new particles.

\section{DISCUSSION}

We are unable to escape the fact that new physics is required to explain the
large scale structure of the universe.  Measurements of the cosmological matter
density, most recently by the WMAP satellite, are precise indications (7\%
error, and fast improving) of {\em something} relating to this new physics.
Identifying exactly what that something is is one of the chief challenges in
physics in the next decade(s).

\begin{acknowledgments}

We thank the organizers of the XXXII SLAC Summer Institute for the invitation
to lecture on dark matter and for putting on such an informative school.  This
work was supported by the U.S. Department of Energy under contract number
\mbox{DE-AC02-76SF00515}.

\end{acknowledgments}


\begin{thebibliography}{99} % Use for 10-99 references

\bibitem{WMAP}
C.~L.~Bennett {\it et al.}  [WMAP Collaboration],
Astrophys.\ J.\ Suppl.\  {\bf 148}, 1 (2003);\\
D.~N.~Spergel {\it et al.}  [WMAP Collaboration],
Astrophys.\ J.\ Suppl.\  {\bf 148}, 175 (2003).

\bibitem{SDSS}
K.~Abazajian {\it et al.}  [SDSS Collaboration],
Astron.\ J.\  {\bf 126}, 2081 (2003);\\
M.~Tegmark {\it et al.}  [SDSS Collaboration],
Phys.\ Rev.\ D {\bf 69}, 103501 (2004).

\bibitem{backgroundlight}
J.~M.~Overduin and P.~S.~Wesson, Phys.\ Rept.\  {\bf 402}, 267 (2004).

\bibitem{cluster_elliptical}
J.~J.~Mohr, A.~E.~Evrard, D.~G.~Fabricant, and M.~J.~Geller, Astrophys.~J. {\bf
447}, 8 (1995).

\bibitem{spergelsteinhardt}
D.~N.~Spergel and P.~J.~Steinhardt,
Phys.\ Rev.\ Lett.\  {\bf 84}, 3760 (2000).

\bibitem{CDM}
P.~J.~E.~Peebles, Astrophys.~J.~Lett.\ {\bf 263}, L1 (1982);\\
G.~R.~Blumenthal, S.~M.~Faber, J.~R.~Primack and M.~J.~Rees,
Nature {\bf 311}, 517 (1984).

\bibitem{WDM}
J.~R.~Bond and A.~S.~Szalay, Astrophys.\ J.\  {\bf 274}, 443 (1983);\\
J.~M.~Bardeen, J.~R.~Bond, N.~Kaiser and A.~S.~Szalay,
Astrophys.\ J.\  {\bf 304}, 15 (1986).

\bibitem{warm_constraint}
V.~K.~Narayanan, D.~N.~Spergel, R.~Dav\'e, C.-P.~Ma,
Astrophys.~J.\ {\bf 543}, L103 (2000).

\bibitem{WDM_bonus}
J.~Sommer-Larsen, A.~Dolgov, Astrophys.~J.\ {\bf 551}, 608 (2001);\\
P.~Bode, J.~Ostriker, N.~Turok, Astrophys.~J.\ {\bf 556}, 93 (2001).

\bibitem{discreteheating}
C.~G.~Lacey and J.~P.~Ostriker, Astrophys.~J.\ {\bf 299}, 633 (1985);\\
B.~Moore, Astrophys.~J.\ {\bf 413}, L93 (1993);\\
H.~Rix and G.~Lake, Astrophys.~J.\ {\bf 417}, L1 (1993).

\bibitem{discretepoisson}
N.~Afshordi, P.~McDonald and D.~N.~Spergel, Astrophys.\ J.\ {\bf 594}, L71
(2003).

\bibitem{microlensing}
C.~Alcock et al., Astrophys.~J.\ {\bf 542}, 281 (2000);\\
C.~Afonso et al., Astron.~Astrophys.\ {\bf 400}, 951 (2003).

\bibitem{fuzzy}
W.~Hu, R.~Barkana and A.~Gruzinov, Phys.\ Rev.\ Lett.\  {\bf 85}, 1158 (2000).

\bibitem{tremainegunn}
S.~Tremaine and J.~E.~Gunn, Phys.\ Rev.\ Lett.\  {\bf 42}, 407 (1979).

\bibitem{KT}
e.g.\ see E.~W.~Kolb, M.~S.~Turner, {\it The Early Universe}, Addison-Wesley
(1990).

\bibitem{gondologelmini}
P.~Gondolo and G.~Gelmini, Nucl.~Phys.\ {\bf B360}, 145 (1991).

\bibitem{darksusy}
P.~Gondolo, J.~Edsjo, P.~Ullio, L.~Bergstrom, M.~Schelke and E.~A.~Baltz,
JCAP {\bf 0407}, 008 (2004).

\bibitem{chaplygin}
A.~Y.~Kamenshchik, U.~Moschella and V.~Pasquier,
Phys.\ Lett.\ B {\bf 511}, 265 (2001);\\
M.~C.~Bento, O.~Bertolami and A.~A.~Sen,
Phys.\ Rev.\ D {\bf 66}, 043507 (2002).

\bibitem{axion}
R.~D.~Peccei and H.~R.~Quinn, Phys.\ Rev.\ Lett.\ {\bf 38},
1440 (1977);\\
S.~Weinberg, Phys.\ Rev.\ Lett.\  {\bf 40}, 223 (1978);\\
F.~Wilczek, Phys.\ Rev.\ Lett.\  {\bf 40}, 279 (1978).

\bibitem{axioncosm}
P.~Sikivie, Phys.\ Rev.\ Lett.\  {\bf 48}, 1156 (1982);\\
J.~Preskill, M.~B.~Wise and F.~Wilczek, Phys.\ Lett.\ B {\bf 120}, 127
(1983);\\
L.~F.~Abbott and P.~Sikivie, Phys.\ Lett.\ B {\bf 120}, 133 (1983);\\
M.~Dine and W.~Fischler, Phys.\ Lett.\ B {\bf 120}, 137 (1983).

\bibitem{axionreview}
M.~S.~Turner, Phys.\ Rept.\ 197, 6797 (1990);\\
G.~G.~Raffelt, Phys.\ Rept.\ 198, 1113 (1990).

\bibitem{sikivie}
P.~Sikivie, Phys.\ Rev.\ Lett.\  {\bf 51}, 1415 (1983).

\bibitem{axionexpt}
For the most recent published results see S.~J.~Asztalos {\it et al.},
Astrophys.\ J.\  {\bf 571}, L27 (2002).

\bibitem{superKneut}
Y.~Fukuda {\it et al.}  [Super-Kamiokande Collaboration],
Phys.\ Rev.\ Lett.\  {\bf 81}, 1562 (1998).

\bibitem{SNO}
Q.~R.~Ahmad {\it et al.}  [SNO Collaboration],
Phys.\ Rev.\ Lett.\  {\bf 89}, 011301 (2002).

\bibitem{neutrinomass}
C.~Weinheimer {\it et al.}, Phys.\ Lett.\ B {\bf 460}, 219 (1999);\\
V.~M.~Lobashev {\it et al.}, Phys.\ Lett.\ B {\bf 460}, 227 (1999).

\bibitem{cowsikmcclelland}
R.~Cowsik and J.~McClelland, Phys.\ Rev.\ Lett.\  {\bf 29}, 669 (1972).

\bibitem{leeweinberg}
B.~W.~Lee and S.~Weinberg, Phys.\ Rev.\ Lett.\  {\bf 39}, 165 (1977).

\bibitem{INTEGRAL}
J.~Knodlseder {\it et al.}, Astron.~Astrophys., in press
(arXiv:astro-ph/0309442);\\
P.~Jean {\it et al.}, Astron.\ Astrophys.\  {\bf 407}, L55 (2003).

\bibitem{MeVDM}
C.~Boehm, D.~Hooper, J.~Silk and M.~Casse,
Phys.\ Rev.\ Lett.\  {\bf 92}, 101301 (2004).

\bibitem{worldsheetSUSY}
P.~Ramond, Phys.\ Rev.\ D {\bf 3}, 2415 (1971).

\bibitem{spacetimeSUSY}
Y.~A.~Golfand and E.~P.~Likhtman, JETP Lett.\  {\bf 13}, 323 (1971);\\
J.~Wess and B.~Zumino, Phys.\ Lett.\ B {\bf 49}, 52 (1974).

\bibitem{MSSM}
H.~P.~Nilles, Phys.\ Rept.\  {\bf 110}, 1 (1984);\\
H.~E.~Haber and G.~L.~Kane, Phys.\ Rept.\  {\bf 117}, 75 (1985).

\bibitem{Rparity}
H.~Pagels and J.~R.~Primack, Phys.\ Rev.\ Lett.\  {\bf 48}, 223 (1982).

\bibitem{gravitino}
J.~R.~Ellis, A.~D.~Linde and D.~V.~Nanopoulos, Phys.\ Lett.\ B {\bf 118}, 59
(1982);\\
T.~Moroi, H.~Murayama and M.~Yamaguchi, Phys.\ Lett.\ B {\bf 303}, 289
(1993).\\
M.~Kawasaki and T.~Moroi, Prog.\ Theor.\ Phys.\  {\bf 93}, 879 (1995).

\bibitem{superwimp}
J.~L.~Feng, A.~Rajaraman and F.~Takayama, Phys.\ Rev.\ Lett.\  {\bf 91}, 011302
(2003);\\
J.~L.~Feng, A.~Rajaraman and F.~Takayama, Phys.\ Rev.\ D {\bf 68}, 063504
(2003).

\bibitem{neutralino}
S.~Weinberg, Phys.\ Rev.\ Lett.\  {\bf 48}, 1303 (1982);\\
H.~Goldberg, Phys.\ Rev.\ Lett.\  {\bf 50}, 1419 (1983);\\
reviewed in G.~Jungman, M.~Kamionkowski and K.~Griest, Phys.\ Rept.\ {\bf 267},
195 (1996).

\bibitem{sneutrino}
J.~S.~Hagelin, G.~L.~Kane and S.~Raby, Nucl.\ Phys.\ B {\bf 241}, 638 (1984);\\
L.~E.~Ibanez, Phys.\ Lett.\ B {\bf 137}, 160 (1984);\\
T.~Falk, K.~A.~Olive and M.~Srednicki, Phys.\ Lett.\ B {\bf 339}, 248 (1994).

\bibitem{axino}
J.~E.~Kim, A.~Masiero and D.~V.~Nanopoulos, Phys.\ Lett.\ B {\bf 139}, 346
(1984);\\
S.~A.~Bonometto, F.~Gabbiani and A.~Masiero, Phys.\ Lett.\ B {\bf 222}, 433
(1989);\\
K.~Rajagopal, M.~S.~Turner and F.~Wilczek, Nucl.\ Phys.\ B {\bf 358}, 447
(1991);\\
L.~Covi, J.~E.~Kim and L.~Roszkowski, Phys.\ Rev.\ Lett.\ {\bf 82}, 4180
(1999).

\bibitem{Qball}
G.~Rosen, J.~Math.~Phys.\ {\bf 9}, 996 (1968);\\
R.~Friedberg, T.~D.~Lee and A.~Sirlin, Phys.\ Rev.\ D {\bf 13}, 2739 (1976);\\
S.~R.~Coleman, Nucl.\ Phys.\ B {\bf 262}, 263 (1985);\\
T.~D.~Lee and Y.~Pang, Phys.\ Rept.\  {\bf 221}, 251 (1992);\\
A.~Kusenko, Phys.\ Lett.\ B {\bf 404}, 285 (1997);\\
A.~Kusenko, Phys.\ Lett.\ B {\bf 405}, 108 (1997);\\
A.~Kusenko and M.~E.~Shaposhnikov, Phys.\ Lett.\ B {\bf 418}, 46 (1998);\\
A.~Kusenko and P.~J.~Steinhardt, Phys.\ Rev.\ Lett.\  {\bf 87}, 141301 (2001).

\bibitem{splitSUSY}
J.~D.~Wells, arXiv:hep-ph/0306127;\\
N.~Arkani-Hamed and S.~Dimopoulos, arXiv:hep-th/0405159.

\bibitem{splitSUSYimp}
G.~F.~Giudice and A.~Romanino, Nucl.\ Phys.\ B {\bf 699}, 65 (2004);\\
N.~Arkani-Hamed, S.~Dimopoulos, G.~F.~Giudice and A.~Romanino,
arXiv:hep-ph/0409232;\\
J.~D.~Wells, arXiv:hep-ph/0411041;\\
A.~Pierce, Phys.\ Rev.\ D {\bf 70}, 075006 (2004).

\bibitem{UED}
I.~Antoniadis, Phys.\ Lett.\ B {\bf 246}, 377 (1990);\\
I.~Antoniadis, K.~Benakli and M.~Quir\'os, Phys.\ Lett.\ B {\bf 331}, 313
(1994);\\
H.~C.~Cheng, K.~T.~Matchev and M.~Schmaltz, Phys.\ Rev.\ D {\bf 66}, 056006
(2002);\\
G.~Servant and T.~M.~P.~Tait, Nucl.\ Phys.\ B {\bf 650}, 391 (2003);\\
E.~W.~Kolb and R.~Slansky, Phys.\ Lett.\ B {\bf 135}, 378 (1984);\\
K.~R.~Dienes, E.~Dudas and T.~Gherghetta, Nucl.\ Phys.\ B {\bf 537}, 47 (1999).

\bibitem{branon}
J.~A.~R.~Cembranos, A.~Dobado and A.~L.~Maroto,
Phys.\ Rev.\ Lett.\  {\bf 90}, 241301 (2003).

\bibitem{mirror}
T.~D.~Lee and C.~N.~Yang, Phys.~Rev.\ {\bf 104}, 254 (1956);\\
I.~Yu.~Kobzarev, L.~B.~Okun, and I.~Ya~Pomeranchuk, Yad.~Fiz.\ {\bf 3}, 1154
(1966);\\
E.~Kolb, D.~Seckel, and M.~Turner, Nature {\bf 314}, 415 (1985);\\
H.~M.~Hodges, Phys.\ Rev.\ D {\bf 47}, 456 (1993);\\
Z.~Berezhiani, D.~Comelli and F.~L.~Villante, Phys.\ Lett.\ B {\bf 503}, 362
(2001);\\
R.~Foot, Acta Phys.\ Polon.\ B {\bf 32}, 2253 (2001).

\bibitem{wimpzilla}
D.~J.~H.~Chung, E.~W.~Kolb and A.~Riotto, Phys.\ Rev.\ Lett.\ {\bf 81}, 4048
(1998);\\
D.~J.~H.~Chung, E.~W.~Kolb and A.~Riotto, Phys.\ Rev.\ D {\bf 59}, 023501
(1999).

\bibitem{PBH}
Ya.~B.~Zel'dovich and I.~D.~Novikov, Sov.~Astron.\ {\bf 10}, 62 (1967);\\
S.~Hawking, Mon.\ Not.\ Roy.\ Astron.\ Soc.\  {\bf 152}, 75 (1971);\\
B.~J.~Carr, Astrophys.\ J.\  {\bf 201}, 1 (1975);\\
K.~Jedamzik, Phys.\ Rev.\ D {\bf 55}, 5871 (1997).

\bibitem{msugra}
A.~H.~Chamseddine, R.~Arnowitt and P.~Nath, Phys.\ Rev.\ Lett.\ {\bf 49}, 970
(1982);\\
R.~Barbieri, S.~Ferrara and C.~A.~Savoy, Phys.\ Lett.\ B {\bf 119}, 343
(1982);\\
N.~Ohta, Prog.\ Theor.\ Phys.\  {\bf 70}, 542 (1983);\\
L.~J.~Hall, J.~Lykken and S.~Weinberg, Phys.\ Rev.\ D {\bf 27}, 2359 (1983).

\bibitem{isajet}
H.~Baer, F.~E.~Paige, S.~D.~Protopescu and X.~Tata, arXiv:hep-ph/0312045.

\bibitem{mcmc}
E.~A.~Baltz and P.~Gondolo, JHEP {\bf 0410}, 052 (2004).

\bibitem{goodmanwitten}
M.~W.~Goodman and E.~Witten, Phys.\ Rev.\ D {\bf 31}, 3059 (1985).

\bibitem{gminus2}
G.~W.~Bennett, {\it et al.}, Phys.~Rev.~Lett.\ {\bf 92}, 161802 (2004);\\
M.~Davier, S.~Eidelman, A.~H\"ocker and Z.~Zhang, Eur.~Phys.~J.~C {\bf 31}, 503
(2003);\\
K.~Hagiwara, A.~D.~Martin, D.~Nomura and T.~Teubner, Phys.\ Rev.\ D {\bf 69},
093003 (2004).

\bibitem{ddcurrent}
A.~Benoit {\it et al.}  [EDELWEISS Collaboration], Phys.\ Lett. B {\bf 545}, 43
(2002);\\
D.~S.~Akerib {\it et al.}  [CDMS Collaboration], Phys.\ Rev.\ Lett.\ {\bf 93},
211301 (2004).

\bibitem{ddproposed}
CDMS II, {\tt http://cdms.berkeley.edu/}\\
CRESST II, {\tt http://wwwvms.mppmu.mpg.de/cresst/}\\
GENIUS, H.~V.~Klapdor-Kleingrothaus~et~al., in ``Beyond the desert 1999,''
(IOP, 2000), p.\ 915;\\
CryoArray, R.~W.~Schnee, D.~S.~Akerib and R.~J.~Gaitskell,
arXiv:astro-ph/0208326;\\
XENON, E.~Aprile {\it et al.}, arXiv:astro-ph/0207670,
{\tt http://www.astro.columbia.edu/\~{}lxe/XENON/}.

\bibitem{dama}
R.~Bernabei {\it et al.}, Riv.\ Nuovo Cim.\  {\bf 26N1}, 1 (2003).

\bibitem{earthsun}
J.~Silk, K.~A.~Olive and M.~Srednicki, Phys.\ Rev.\ Lett.\ {\bf 55}, 257
(1985);\\
K.~Freese, Phys.\ Lett.\ B {\bf 167}, 295 (1986);\\
L.~M.~Krauss, M.~Srednicki and F.~Wilcek, Phys.\ Rev.\ D {\bf 33}, 2079 (1986).

\bibitem{oldMOND}
M.~Milgrom, Astrophys.\ J.\  {\bf 270}, 365 (1983);\\
M.~Milgrom, Astrophys.\ J.\  {\bf 270}, 371 (1983);\\
M.~Milgrom, Astrophys.\ J.\  {\bf 270}, 384 (1983).

\bibitem{relMOND}
J.~Bekenstein and M.~Milgrom 1984, Astrophys.~J.\  {\bf 286}, 7 (1984);\\
J.~D.~Bekenstein, Phys.\ Lett.\ B {\bf 202}, 497 (1988);\\
M.~E.~Soussa and R.~P.~Woodard, Class.\ Quant.\ Grav.\  {\bf 20}, 2737 (2003).

\bibitem{newMOND}
J.~D.~Bekenstein, Phys.\ Rev.\ D {\bf 70}, 083509 (2004).

\end{thebibliography}
\end{document}